\documentclass{article} 
\usepackage{iclr2025_conference,times}

\usepackage{amsmath,amsfonts,bm}

\def\eqref#1{equation~\ref{#1}}

\def\1{\bm{1}}

\DeclareMathAlphabet{\mathsfit}{\encodingdefault}{\sfdefault}{m}{sl}
\SetMathAlphabet{\mathsfit}{bold}{\encodingdefault}{\sfdefault}{bx}{n}

\usepackage{hyperref}
\usepackage{url}
\usepackage{graphicx}
\usepackage{subcaption}
\usepackage{tcolorbox}
\usepackage[normalem]{ulem}
\usepackage{xcolor}
\usepackage{listings}
\title{Generating Structured Plan Representation of Procedures with LLMs}

\author{
Deepeka Garg\textsuperscript{1}\thanks{deepeka.garg@jpmorgan.com}, 
Sihan Zeng\textsuperscript{2}, 
Sumitra Ganesh\textsuperscript{3}, 
Leo Ardon\textsuperscript{1} \\
\textsuperscript{1}JP Morgan AI Research, London, UK \\
\textsuperscript{2}JP Morgan AI Research, Palo Alto, USA \\
\textsuperscript{3}JP Morgan AI Research, New York, USA \\
}

\iclrfinalcopy 
\begin{document}

\maketitle

\begin{abstract}
In this paper, we address the challenges of managing Standard Operating Procedures (SOPs), which often suffer from inconsistencies in language, format, and execution, leading to operational inefficiencies. Traditional process modeling demands significant manual effort, domain expertise, and familiarity with complex languages like Business Process Modeling Notation (BPMN), creating barriers for non-techincal users. We introduce SOP Structuring (SOPStruct), a novel approach that leverages Large Language Models (LLMs) to transform SOPs into decision-tree-based structured representations. SOPStruct produces a standardized representation of SOPs across different domains, reduces cognitive load, and improves user comprehension by effectively capturing task dependencies and ensuring sequential integrity. Our approach enables leveraging the structured information to automate workflows as well as empower the human users. By organizing procedures into logical graphs, SOPStruct facilitates backtracking and error correction, offering a scalable solution for process optimization. We employ a novel evaluation framework, combining deterministic methods with the Planning Domain Definition Language (PDDL) to verify graph soundness, and non-deterministic assessment by an LLM to ensure completeness. We empirically validate the robustness of our LLM-based structured SOP representation methodology across SOPs from different domains and varying levels of complexity. Despite the current lack of automation readiness in many organizations, our research highlights the transformative potential of LLMs to streamline process modeling, paving the way for future advancements in automated procedure optimization.   
\end{abstract}

\section{Introduction}
\label{sec.introduction}
Standard Operating Procedures (SOP) are essential guidelines that provide detailed step-by-step instructions to execute critical daily operations in various disciplines. They specify what actions to take, how to perform them, and when to execute them, ensuring that operations are carried out with precision and consistency, leading to reliable outcomes. Real-life procedures are often complex and consist of multiple interrelated instructions. These procedures often involve long-horizon planning, requiring multiple interconnected actions that span extended time frames to achieve specific objectives \cite{safa2024systematic}. This type of sequential task planning presents unique challenges; the execution of actions at one point in time can significantly influence subsequent actions and outcomes. Managing these temporal dependencies and addressing the combinatorial complexity of such tasks makes long-horizon planning particularly difficult.

\begin{figure}[h]
    \begin{center}
      \includegraphics[width=0.8\textwidth]{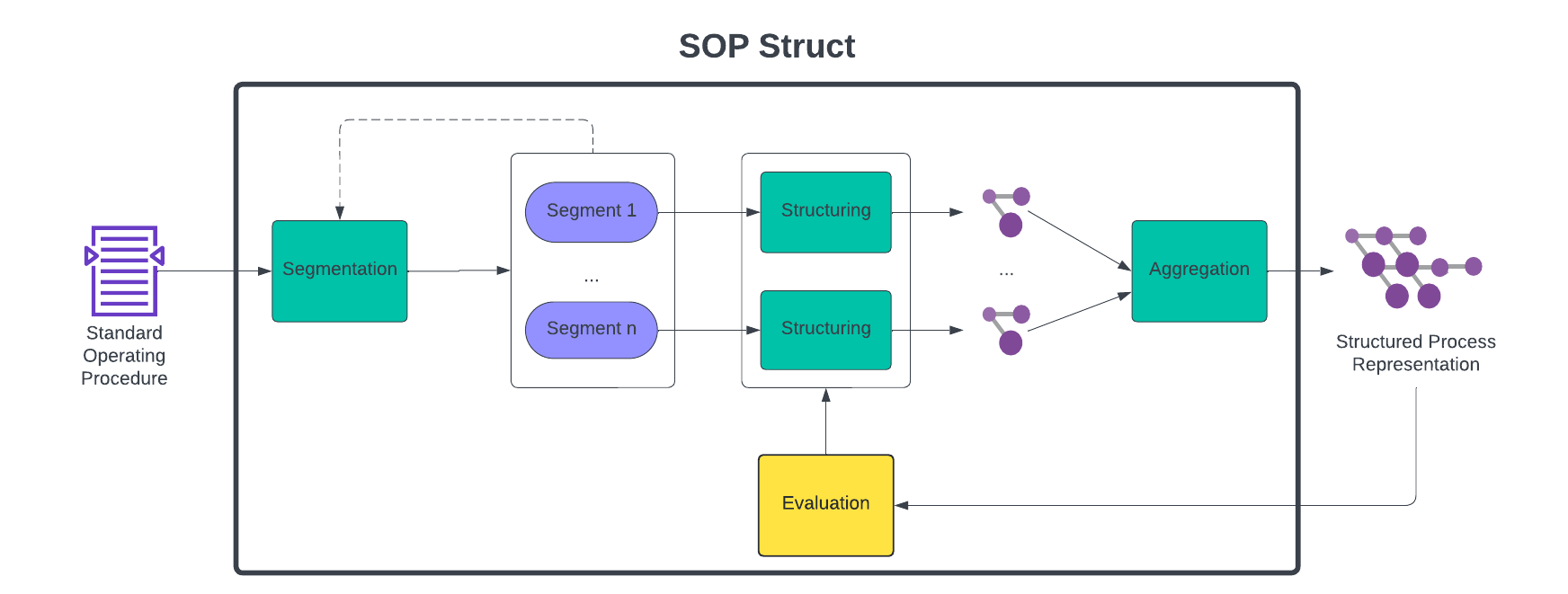}
       
    \end{center}
    \caption{Our Structured Process Generation Methodology: SOPStruct}
    \label{fig:methodology}
\end{figure}

SOPs are often mandated by industry regulations, making them legally binding. Well-structured SOPs not only ensure compliance but also promote sound business practices \cite{gough2009standard}. Although business standards exist \cite{aguilar2004business}, the structure and documentation of SOPs are largely determined by Subject Matter Experts (SMEs), often expert in their line of business but missing the technical acumen and the time to learn and implement complex formal languages for SOPs, which are not always aligned with their business needs. The reliance on human interpretation without fixed templates or syntax not only increases the risk of human error but can leave human users feeling overwhelmed, especially if the procedure is lengthy and complex. Moreover, it also hampers the AI-based process automation or facilitation of hybrid human-AI execution architectures. 

To address this issue, two potential approaches exist; standardizing SOP creation from the outset or enhancing the comprehensibility of existing SOPs. The former involves establishing uniform templates and guidelines, which can be resource-intensive and require widespread changes. The latter, which is the focus of this paper, aims to improve the clarity and usability of existing SOPs. By creating logical interpretations of these procedures, we transform them into a structured format without necessitating a complete overhaul of current practices. This practical approach can enhance the utility of existing SOPs and facilitate their integration with AI-driven solutions.

Given the natural language handling and reasoning capabilities of LLMs, in this paper, we tackle the SOP structuring challenge by leveraging LLMs to convert unstructured natural language SOPs into a structured Directed Acyclic Graph (DAG) format, which serves as representations of task workflows, capturing both logical and temporal dependencies. Our method breaks down lengthy, unstructured SOPs into subtasks and captures the dependencies between these steps. This structured representation simplifies the SOP understanding and makes it more amenable to automatic processing, improving the workflow efficiency.

A key innovation of our approach is the deterministic evaluation of structured SOPs (DAGs) using a PDDL-based planner, which provides a scalable and objective method for assessing the logical soundness and connectivity of the generated task plans. This contrasts with standard process representation BPMN literature \cite{kopke2024introducing}, which often relies on human evaluators, posing scalability challenges, and with modern LLM-based evaluation methods \cite{tang2023toolalpaca}, where confidence in accuracy can be limited due to the inherent uncertainties of LLM outputs.

We believe this work opens avenues for a comparative analysis of different process planning approaches, aiming to unify logical plan generation across diverse domains. 
Experimental results show that our method effectively models task dependencies and generalizes across domains (such as BPMN-type business tasks and non-business tasks), highlighting its adaptability and broad applicability.

\section{Related Work}
\label{sec.relatedwork}
The management and automation of Standard Operating Procedures (SOPs) have long been a challenge due to their unstructured nature and reliance on human expertise. Traditional process modeling approaches, such as Business Process Modeling Notation (BPMN) and Planning Domain Definition Language (PDDL), require significant manual effort and domain knowledge to encode task-specific rules and dependencies \cite{fox2003pddl2}. While effective in structured environments, these methods often lack the flexibility needed to handle the diverse and dynamic nature of real-world SOPs.

Recent advancements in Large Language Models (LLMs) have opened new avenues for task planning and structuring. LLMs have been employed in various domains to generate and optimize workflows, demonstrating their potential to handle complex reasoning tasks \cite{sharma2021skill, huang2022language, song2023llm, singh2023progprompt}. However, much of the existing research focuses on prompting strategies, such as ``Chain of Thought'' and ``Tree of Thoughts'', which aim to enhance reasoning capabilities by modeling thought processes as linear or hierarchical structures \cite{wei2022chain, yao2023react, hong2023metagpt, luo2023obtaining, yao2024tree}. These strategies have proven effective in guiding LLMs through structured reasoning paths, yet they focus on the reasoning process itself rather than the organization of complex procedural information.

\begin{figure}[t]
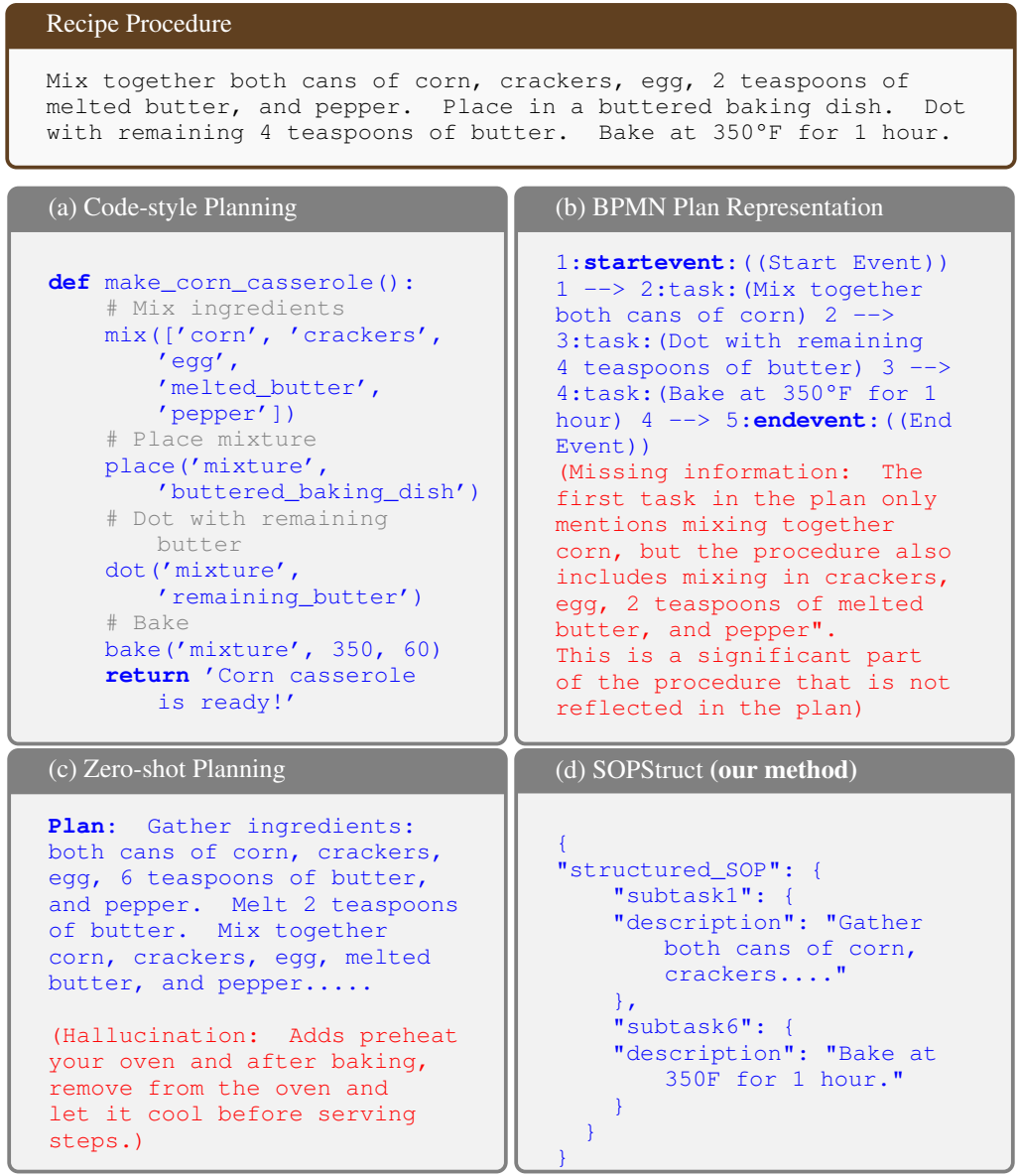

\centering
\begin{tcolorbox}[colframe=brown!50!black, colback=gray!5!white, title=Recipe Procedure, width=0.97\textwidth, fontupper=\ttfamily\small]
Mix together both cans of corn, crackers, egg, 2 teaspoons of melted butter, and pepper. Place in a buttered baking dish. Dot with remaining 4 teaspoons of butter. Bake at 350°F for 1 hour.
\end{tcolorbox}
\begin{subfigure}[b]{0.48\textwidth}
\begin{tcolorbox}[title= (a) Code-style Planning, colframe=black!50!white, colback=white!90!gray, height=7.5cm]

\begin{lstlisting}
def make_corn_casserole():
    # Mix ingredients
    mix(['corn', 'crackers', 'egg', 'melted_butter', 'pepper'])
    # Place mixture
    place('mixture', 'buttered_baking_dish')
    # Dot with remaining butter
    dot('mixture', 'remaining_butter')
    # Bake 
    bake('mixture', 350, 60)
    return 'Corn casserole is ready!'
\end{lstlisting}
\end{tcolorbox}
\end{subfigure}
\begin{subfigure}[b]{0.48\textwidth}
\begin{tcolorbox}[title= (b) BPMN Plan Representation, colframe=black!50!white, colback=white!90!gray, fontupper=\ttfamily\small, height=7.5cm]
\textcolor{blue}{1:\textbf{startevent}:((Start Event))
1 --> 2:task:(Mix together both cans of corn)
2 --> 3:task:(Dot with remaining 4 teaspoons of butter)
3 --> 4:task:(Bake at 350°F for 1 hour)
4 --> 5:\textbf{endevent}:((End Event))} \\ 
\textcolor{red}{(Missing information: The first task in the plan only mentions mixing together corn, but the procedure also includes mixing in crackers, egg, 2 teaspoons of melted butter, and pepper". \\ This is a significant part of the procedure that is not reflected in the plan)}
\end{tcolorbox}

\end{subfigure}
\vspace{0.5cm}
\begin{subfigure}[b]{0.48\textwidth}
\begin{tcolorbox}[title= (c) Zero-shot Planning, colframe=black!50!white, colback=white!90!gray, fontupper=\ttfamily\small, height=5.7cm]
\textcolor{blue}{\textbf{Plan}: Gather ingredients: both cans of corn, crackers, egg, 6 teaspoons of butter, and pepper. Melt 2 teaspoons of butter. Mix together corn, crackers, egg, melted butter, and pepper.....}\\
\\
\textcolor{red}{(Hallucination: Adds preheat your oven and after baking, remove from the oven and let it cool before serving steps.)}
\end{tcolorbox}
\end{subfigure}
\begin{subfigure}[b]{0.48\textwidth}
\begin{tcolorbox}[title= (d) SOPStruct \textbf{(our method)}, colframe=black!50!white, colback=white!90!gray, fontupper=\ttfamily\small, height=5.7cm]
\begin{lstlisting}
{
"structured_SOP": {
    "subtask1": {
    "description": "Gather both cans of corn, crackers...."
    },
    "subtask6": {
    "description": "Bake at 350F for 1 hour."
    }
  }
}
\end{lstlisting}
\end{tcolorbox}
\end{subfigure}

\caption{Comparison of plans generated by different methods for a recipe procedure, (a) Code-style, (b) BPMN, (c), Zero-shot, and (d) SOPStruct. For brevity, we only include the starting and final subtasks with descriptions for SOPStruct.}
\label{fig:method-comparison}
\end{figure}

In contrast, our work introduces a novel \emph{structuring} method that leverages LLMs to transform unstructured SOPs into DAGs, emphasizing the structuring of information rather than the reasoning process alone. This approach captures both logical and temporal dependencies within task workflows, providing a clear and interpretable representation of complex procedures. By focusing on the organization of SOPs into graph-based structures, we address the need for a scalable and adaptable solution that enhances workflow efficiency. 

       

Furthermore, our method addresses the limitations of existing evaluation techniques by employing a dual framework that combines deterministic assessments using PDDL with non-deterministic evaluations by LLMs. This ensures the logical soundness and completeness of the generated DAGs, contrasting with BPMN literature that often relies on human evaluators \cite{kopke2024introducing}.

Recently, self-reflection and self-correction mechanisms have been explored as methods to enhance the performance of LLMs in generating accurate and coherent outputs \cite{shinn2023reflexion, madaan2024self}. These approaches involve iterative processes where the model evaluates and refines its own outputs, aiming to improve reasoning and decision-making capabilities. However, such mechanisms can introduce additional complexity and computational overhead. Our work posits that with carefully designed prompts, LLMs can be guided to organize and reason about information effectively from the outset. By structuring prompts to align with the desired output format, such as organizing procedures into DAGs, we can leverage the inherent capabilities of LLMs to produce high-quality outputs without the need for iterative self-reflection or correction. This approach not only simplifies the process but also demonstrates the potential of LLMs to achieve robust performance through strategic prompting, reducing reliance on post-hoc adjustments and enhancing efficiency in handling complex procedural tasks.

\section{Background}
\label{sec.background}
This section provides the necessary background for understanding our approach.
\subsection{Structured Representation of Procedures}

Standard Operating Procedures serve as a means to maintain and transfer knowledge between people, acting as the golden source of information on how to execute procedures. They are used across various tasks and domains ranging from straightforward daily tasks, such as booking a bus ticket, to more intricate operations, such as understanding the requirements for working and living in a specific country. By nature, SOPs provide a detailed description of the logical sequence of actions required to achieve specific objectives. Although they are typically written in plain language with some flexibility in style and format, they must follow some logical structure for the readers to understand and execute.

We focus on the logical flow described in the SOP that details the procedure to follow, abstracting away the aspects such as language and formatting. The logical structure of an SOP is formalized as a graph, where the steps (or sub-tasks) are represented as vertices and the edges capture both logical and temporal dependency relationships between them.

Formally, an SOP is associated with a \emph{logical} graph $G = (V, E)$, where $V$ is the set of vertices representing the sub-tasks of the SOP and where $E$ is the set of directed edges, where an edge  $(v_i, v_j) \in E $ indicates that the sub-task $v_j$ depends on $v_i$. Following this structure, if there is a directed edge from node $v_i$ to node $v_j$, then $v_i$ is considered a \emph{dependency} of $v_j$. The parent node $v_i$ may produce an output required as an input for the child node $v_j$ or it could simply indicate that $v_i$ should be executed before $v_j$. This parent-child relationship captures the logical and temporal dependencies necessary for the execution of complex tasks. Using the graph representation of the SOP, we say a procedure is \emph{complex} if its associated graph is complex, where the graph complexity can be loosely associated with the graph's depth and/or the number of branches.

This graphical representation helps uncover the logical dependency structure described in the SOP, making it more amenable to processing, visualization and optimization. 

\subsection{Classical Planning}

One of the key contributions of our work is the application of a classical planning methodology to evaluate the structured SOP generated, thereby providing some formal guarantees on the validity of the solutions produced. The Planning Domain Definition Language (PDDL) is a formal language used to describe planning problems and domains in artificial intelligence \cite {aeronautiques1998pddl}. It offers a standardized framework for defining the initial state, the goal state, and the actions that can be performed to transition between states. 
%
PDDL-based planner operates by taking as input a domain description and a problem description. The domain description specifies the types of objects, predicates, and actions available, including their preconditions and effects defined using predicates. The problem description defines the specific initial state of the world and the desired goal state. The planner's task is to generate a sequence of actions, or a plan, that transforms the initial state into a state that satisfies the goal conditions.
These planners systematically explore the state space delineated by the problem, employing search algorithms to identify optimal or feasible solutions, while ensuring the correctness and validity of the generated plans.

The proposed graphical representation of the SOP makes classical planning particularly well-suited for evaluating the graph's structure. Specifically, the problem can be conceptualized as traversing the graph representing the SOP as a human would do, wherein, starting from an initial state (the root of the structured SOP), the objective is to reach one of the goal states (the leaf nodes) by navigating through the edges that encode the dependency structure of the SOP. 

\section{Methodology}
\label{sec.ourmethod}
In this paper, we introduce SOPStruct (SOP Structuring Agent) to create structured representation of complex, long-horizon decision-making processes. Unlike traditional methods that rely on predefined planning languages such as PDDL \cite{fox2003pddl2} or frameworks like (Hierarchical Task Network) HTNs \cite {georgievski2014overview}, our method dynamically constructs structured representations in the form of Directed Acyclic Graphs (DAGs). These graphs effectively capture both the logical and temporal dependencies inherent in the planning process. Real-world procedures typically encompass a mix of sequential and concurrently occurring tasks. For example, in a bicycle order supply SOP, once an order is accepted, the procurement of bicycle parts and their assembly (depending on existing inventory) can proceed simultaneously. A DAG can effectively capture these types of sequential and parallel relationships between events, providing a clear and structured process flow. SOPStruct comprises of three primary phases (shown in Figure ~\ref{fig:methodology}): SOP Segmentation, SOP Structure Generation and Evaluation, detailed below.

\subsection{SOP Segmentation Methodology}
The segmentation phase begins with an LLM (GPT4 in our case) analyzing the SOP document to identify distinct process segments. Using its natural language understanding, the LLM detects context shifts and boundaries between process steps, pinpointing the start and end of each segment. Subsequently, we programmatically extract the text within these LLM-identified boundaries to prepare each segment for structured representation.

In this phase, we partition the procedure \( P \) into smaller, coherent segments \( \{S_k\}_{k=1}^m \). Each segment \( S_k \) is a manageable subset of the overall procedure, allowing for accurate transformation into a subgraph of the DAG. This segmentation step ensures that the integrity and completeness of information are maintained, facilitating the construction of a comprehensive DAG \( G = (V, E) \) without requiring an overwhelming amount of computation or losing critical task details. By segmenting the procedure, we ensure that the resulting DAG accurately reflects the logical and temporal dependencies of the entire SOP, while maintaining scalability and reliability in the representation process.

Without segmentation, directly generating a structured representation leads to the loss of fine-grained details, capturing only high-level information. Although LLMs with large token limits (e.g., GPT-4 with a 32K token context) can process lengthy SOPs, they still struggle to capture intricate dependencies and nuances inherent in the procedures. Thus, even techniques that extend context length, such as RoPE \cite{li2024extending}, do not address this loss of granular detail. By breaking SOPs into manageable segments, we capture even the most subtle aspects of the process, preventing the omission of crucial process-specific information in the final DAG. Moreover, segmentation can be applied recursively for finer decomposition, ensuring both segment completeness and manageability.

In conclusion, this segmentation phase guarantees consistent structured representation quality across SOPs of varying lengths.

\subsection{SOP Structure Generation Methodology}
In the structure generation phase, our approach leverages the LLM to decompose the SOP segments into a series of subtasks, each representing a vertex of the DAG. Dependencies between these subtasks are captured as edges between vertices, allowing for topological sorting and ensuring that the SOP can be executed in the correct order. We adhere to traditional DAG conventions by specifying names, descriptions, dependencies and output for each node, while extending this formalism to include additional attributes such as inputs from dependencies. This specifies which outputs from previous nodes are used as inputs in the current node, resulting in a clearly defined graph connectivity. Furthermore, each node is assigned a category attribute to identify its type: whether it is a decision step, an action to execute, or domain-specific knowledge. This additional information provides a richer encoding of the SOP, enhancing user understanding and facilitating downstream automated execution.

For each segment \( S_k \), we generate a set of subtasks \( \text{ST}(S_k) = \{s_i\}_{i=1}^{n_k} \), where each subtask \( s_i \) corresponds to a node in the DAG. Each sub-task is a node in $V$ and is defined with the following information:

\textbf{Name}: The name of the subtask.

\textbf{Description}: A detailed process description of the current subtask.

\textbf{Dependencies}: A list of other subtasks on which this subtask depends (i.e., its parent nodes).

\textbf{Inputs}: Inputs required for the subtask that comes from the initial state and not from any dependency subtask.

\textbf{Inputs from Dependencies}: A mapping of inputs received from dependency subtasks.

\textbf{Output}: A list of outputs produced by the subtask.

\textbf{Category}: This field specifies the operational nature of the subtask, categorized into ``\textit{Human Input}'' (receiving and saving user-provided information), ``\textit{Information Processing}'' (analyzing, verifying, or manipulating data), ``\textit{Information Extraction}'' (actively searching for information that is not explicitly provided in the SOP), ``\textit{Knowledge}'' (stating general background information that is not directly actionable but can provide additional context) and ``\textit{Decision}'' (stating decisions, judgments interpretations or conclusions). 

This comprehensive specification ensures that each subtask is clearly defined and contextualized. By structuring each subtask as a JSON object; a format that aligns with LLMs' strength in generating structured outputs that maintain key-value relationships, adhere to strict syntax rules, and conform to schema constraints \cite{liu2024we}, we provide the model with a detailed schema that includes attributes such as name, dependencies, inputs, and outputs etc. This multi-dimensional constrained specification enables the LLM to reason about these attributes from the outset while mitigating well-known hallucination issues \cite{maynez2020faithfulness}, enabling it to correctly decompose complex, overwhelming procedures into manageable subtasks. 

\subsection{Evaluation Methodology} \label{sec:evaluation}

We leverage both deterministic and non-deterministic approaches to assess the quality and completeness of the DAG generated by our method. Our evaluation methodology is novel and supports cases where the ground truth is not available, which is common for practical problems. We define several key metrics to evaluate the DAG, each addressing a different aspect of its validity and utility.

\noindent\textbf{Structured Plan Score.} This metric assesses whether the graph  $(G = (V, E))$ is connected, ensuring that traversal is possible from the initial node $(s_{\text{start}})$ to the final node $(s_{\text{end}})$ based on the generated dependency structure. This is evaluated deterministically using a classical PDDL planner. 

\noindent\textbf{Dependency Score.} For each subtask $s_i\in V$, this metric ensures that it only expects data from subtasks explicitly listed in its dependencies $D(s_i)=\{v\in V: (v,s_i)\in E\}$. The validation fails if an input is expected from a subtask not present in $D(s_i)$. This is a deterministic evaluation.

\noindent\textbf{Input from Dependency Score.} This metric checks that input data received from a dependency node matches the output of that dependency node. For each subtask $(s_i)$, the required inputs must map to the outputs of other subtasks. This is evaluated deterministically.

\noindent\textbf{Plan Initial State Validation Score.} This non-deterministic metric evaluates whether the graph accurately reflects the input information specified in the procedure. We use a language model to compare the initial state extracted from the graph with the initial state specified in the instructions, accounting for semantic variations. The initial state in the DAG as the union of the ``Inputs'' attribute of the nodes.  

\noindent\textbf{Plan Goal State Validation Score.} Similar to the initial state alignment, this non-deterministic metric assesses whether the graph accurately reflects the goal (output) information specified in the SOP. A language model is used to compare the goal state from the graph with the goal state from the instructions. 
The goal in the DAG is defined as the set of outputs produced by subtasks that are not consumed as inputs by any subsequent (child) subtasks. 

\noindent\textbf{Plan Completeness Score.} This metric checks if the graph is complete and encodes all the relevant information from the SOP. We prompt a language model to ensure that no critical details are overlooked, providing an additional layer of assurance.

Note: For baselines, deterministic evaluation is not feasible, therefore, we use a language model for these assessments.


\subsection{Classical Planning Approach for Graph Validation}

To test the connectivity and dependency structure of the graph, measured by the Structured Plan Score, Dependency Score and Input from Dependency Score, we employ a planner that models the problem as a meta-planning problem with the domain defined as follow:

\textbf{Predicates}
\begin{itemize}
    \item[] \texttt{available: ?v-variable}: Indicates whether the variable \texttt{v} is available for use.
    \item[] \texttt{required-input: ?v-variable}: Indicates that the variable \texttt{v} must be available for a subtask to be executed.
    \item[] \texttt{required-input: ?v-variable, ?s-subtask}: Indicates that the variable \texttt{v} must be available for the subtask \texttt{s} to be executed.
    \item[] \texttt{subtask-output: ?v-variable, ?s-subtask}: Indicates that the variable \texttt{v} will be made available once the subtask \texttt{s} is executed.
    \item[] \texttt{map: ?v1-variable, ?v2-variable}: Indicates that the variable with name \texttt{v1} can be mapped with the variable with name \texttt{v2}.
\end{itemize}

\textbf{Actions}
\begin{itemize}
    \item[] \texttt{execute-subtask: ?s-subtask}: Execute the subtask \texttt{s} if the required inputs are available, making the outputs of \texttt{s} available.
    \item[] \texttt{assign: ?v1-variable, ?v2-variable}: Assigns the truth value of \texttt{v1} to \texttt{v2} if a mapping between \texttt{v1} and \texttt{v2} exist.
\end{itemize}

For each graph, we automatically generate the problem description:

\textbf{Objects}: All subtasks' name, inputs and outputs used to solve the problem.

\textbf{Initial State}: Input variables not coming from dependencies' outputs, and assumed to compose the initial state.

\textbf{Required Inputs}: Subtask's inputs required to execute the subtask.

\textbf{Subtask Effects}: Output variables made available once a subtask is executed.

\textbf{Goal}: Outputs of all subtasks, with additional checks to ensure alignment with the specified goals.

We generate a new problem definition for each DAG and use a PDDL-based planner to search a plan that can traverse the graph produce by SOPStruct. If a plan is found the graph is guaranteed to be sound with a well structured dependency graph between the subtasks.

\section{Experiments}
\label{sec:experiments}

\subsection{Datasets}
To build and evaluate our methodology, we selected three datasets, each representing varying levels of complexity and types of procedures. This dataset selection aims to test the flexibility and effectiveness of our approach across a spectrum of complexities, from simpler and shorter to more challenging and longer procedures. 

\begin{itemize}
    \item \textbf{Nestful API Dataset} \cite{basu2024nestful}: Representing the lower end of the complexity spectrum, this dataset includes procedural instructions involving nested API calls. It serves as a preliminary test to assess basic procedural understanding and sequence management.
    
    \item \textbf{Recipe Dataset (RecipeNLG)} \cite{bien2020recipenlg}: Positioned at medium complexity, this dataset challenges the conversion of culinary procedure instructions into structured representations. It tests the ability to handle semi-structured data while maintaining coherence in procedure representation.
    
    \item \textbf{Business Process Dataset} \cite{monti2024nl2processops}: This most complex dataset includes intricate textual descriptions of business processes. It is crucial for evaluating how effectively our methodology and the baseline methods capture and model intricate, multi-step business operations and their dependencies, consequently testing the limits of what each approach can handle.
\end{itemize}



\begin{table}[!ht]
\caption{Evaluation Results (\%) on Nestful API Dataset \cite{basu2024nestful}}
\label{metrics-comparison-nestful-table}
\begin{center}
\begin{tabular}{p{4.4cm}p{1.85cm}p{1.85cm}p{1.85cm}p{1.9cm}}
\hline
\textbf{Metric} & \textbf{Zero-Shot Generation} & \textbf{Code-Style Generation} & \textbf{BPMN \newline Generation} & \textbf{Our Method \newline (SOPStruct)} \\
\hline
Structured Plan Score & 66 & 89.65 & 84 & 100 \\
Plan Initial State Validation & 52.17 & 94.34 & 85.67 & 95.65 \\
Plan Goal State Validation & 51.09 & 83.48 & 89.57 & 97.82 \\
Plan Completeness Score & 50.0 & 72.83 & 78.47 & 95.65 \\
Dependency Score & N/A & N/A & N/A & 100 \\
Inputs from Dependency Score & N/A & N/A & N/A & 100 \\
\hline
\end{tabular}
\end{center}
\end{table}

\begin{table}[!ht]
\caption{Evaluation Results (\%) on RecipeNLG Dataset \cite{bien2020recipenlg}}
\label{metrics-comparison-recipe-table}
\begin{center}
\begin{tabular}{p{4.4cm}p{1.85cm}p{1.85cm}p{1.85cm}p{1.9cm}}
\hline
\textbf{Metric} & \textbf{Zero-Shot Generation} & \textbf{Code-Style Generation} & \textbf{BPMN \newline Generation} & \textbf{Our Method \newline (SOPStruct)} \\
\hline
Structured Plan Score & 73.39 & 90.42 & 76.35 & 100 \\
Plan Initial State Validation & 94.35 & 92.95 & 82.50 & 96.66 \\
Plan Goal State Validation & 77.17 & 90.13 & 79.46 & 93.33 \\
Plan Completeness Score & 59.13 & 89.24 & 78.15 & 92.52 \\
Dependency Score & N/A & N/A & N/A & 100 \\
Inputs from Dependency Score & N/A & N/A & N/A & 100 \\
\hline
\end{tabular}
\end{center}
\end{table}

\begin{table}[!ht]
\caption{Evaluation Results (\%) on Business Process Dataset \cite{monti2024nl2processops}}
\label{metrics-comparison-business-processtable}
\begin{center}
\begin{tabular}{p{4.4cm}p{1.85cm}p{1.85cm}p{1.85cm}p{1.9cm}}
\hline
\textbf{Metric} & \textbf{Zero-Shot Generation} & \textbf{Code-Style Generation} & \textbf{BPMN \newline Generation} & \textbf{Our Method \newline (SOPStruct)} \\
\hline
Structured Plan Score & 80.78 & 66.17 & 62.19 & 100 \\
Plan Initial State Validation & 92.17 & 90.48 & 92.11 & 95.65 \\
Plan Goal State Validation & 86.09 & 87.39 & 86.71 & 95.65 \\
Plan Completeness Score & 71.74 & 55.65 & 52.31 & 94 \\
Dependency Score & N/A & N/A & N/A & 100 \\
Inputs from Dependency Score & N/A & N/A & N/A & 100 \\
\hline
\end{tabular}
\end{center}
\end{table}

\subsection{Baselines}
To systematically evaluate our  methodology, we have selected a diverse and representative set of baseline methods. These baselines cover some of the key approaches used in natural language task and procedure planning. To ensure consistency in output generation, we use the GPT-4 model for both our method and the baselines. Baselines include: (a) \textbf{Standard zero shot generation} \cite{huang2022language} directly utilizes the inherent reasoning capabilities of LLMs to generate procedure representations. Unlike our SOPStruct and other baseline methods that rely on structured prompts or in-context examples as stimuli for structured procedure generation, zero-shot planning generates output representations without any external stimulus. 
(b) \textbf{Code style prompt}; PROGPROMPT \cite{singh2023progprompt} employs program-style specifications of environment objects and actions. (c) \textbf{LLM-based Process Modeling Approach}: BPMN (Business Process Model Notation) \cite{monti2024nl2processops} leverages LLMs to generate BPMN diagrams from natural language procedure inputs.


\subsection{Evaluation } 

We assess the effectiveness of our methodology compared to the established baselines on the metrics introduced in Section~\ref{sec:evaluation}, namely, structured plan score, plan initial state validation score, plan goal state validation score, plan completeness score, dependency score, and inputs from dependency score. Note that the last two are graph-specific metrics that do not apply to the baseline methods. To ensure consistency in assessment, all non-deterministic evaluations for both our method and the baselines are conducted using the GPT-4 model. 

\section{Results and Analysis}
\label{sec.resultsandanalysis}
We present our empirical findings in Tables~\ref{metrics-comparison-nestful-table}, \ref{metrics-comparison-recipe-table}, and \ref{metrics-comparison-business-processtable}. SOPStruct demonstrates the most consistent performance across all datasets, significantly outperforming the baselines. Its structured, graph-based representation, derived from procedural segments, effectively captures logical and temporal dependencies, enabling it to handle procedures of varying complexity. Zero-shot planning operates in an open-ended manner with minimal constraints, as it does not rely on predefined structures or in-context examples. This lack of constraints can lead to challenges such as hallucinations, as shown in ~\ref{fig:method-comparison} (c). For example, in a recipe task, the LLM introduces preheating and serving instructions into the plan, which are not present in the original procedure. 

Both BPMN and code-style methods exhibit fewer procedural hallucinations compared to zero-shot planning. Code-style and BPMN plan generation provide improved adherence to constraints,  leading to the generation of more reliable plans. Among the two, code-style plans outperform BPMN, which often misses adding critical procedural details to the generated plans. For instance, as shown in ~\ref{fig:method-comparison} (b), the BPMN plan omits key steps like mixing in crackers, egg, melted butter, and pepper. Code-style plans outperformed BPMN by better capturing dependencies and logical transitions (as seen in ~\ref{fig:method-comparison} (a)). However, the code-style method underperformed compared to SOPStruct, particularly in handling longer and more complex business procedures. Code-style plans tend to omit granular yet important details. For instance, in an accident reporting business procedure, the code-style plan fails to include specific aid organizations, such as the volunteer fire brigade and the volunteer water rescue service, which are explicitly mentioned in the original procedure. In contrast, SOPStruct generates a complete and detailed structured representation, preserving all essential procedural elements. We hypothesize that segmenting the SOP into manageable parts improves its structured representation.

Contrary to expectations based on dataset complexity, Zero-shot planning performs worst on the Nestful API dataset, which is the simplest dataset in terms of procedural complexity. We hypothesize that this is due to its open-ended nature that increases the likelihood of hallucinations. For more constrained datasets like RecipeNLG and Business Processes, where procedure descriptions inherently limit the scope for inference, Zero-shot planning shows comparatively better performance. This highlights the sensitivity of Zero-shot planning to dataset structure and the importance of constraints in mitigating hallucinations.




\section{Conclusion}
We introduce an efficient and scalable approach, SOPStruct, that leverages LLMs to transform unstructured procedural information into graph-based structured formats, enabling interpretable and standardized representations SOPs across different domains. Our DAG-based representation captures logical and temporal dependencies between the procedure steps while ensuring an acyclic execution flow for reliable procedure completion. The scalability and robustness of our method make it well-suited for large-scale SOP management in real-world applications. In practice, defining a single ground truth for SOPs is often infeasible, as procedures can be structured in multiple valid ways, varying in granularity. To address this challenge, we introduce a novel dual evaluation paradigm that combines PDDL-based planning with LLM-based assessments, enabling structured and scalable evaluation of DAG quality. This work lays the groundwork for future advancements in procedural automation, domain-specific SOP optimization, and large-scale workflow efficiency.

\section{Disclaimer}
This paper was prepared for informational purposes by the Artificial Intelligence Research group of JPMorgan Chase \& Co and its affiliates (“J.P. Morgan”) and is not a product of the Research Department of J.P. Morgan.  J.P. Morgan makes no representation and warranty whatsoever and disclaims all liability, for the completeness, accuracy or reliability of the information contained herein.  This document is not intended as investment research or investment advice, or a recommendation, offer or solicitation for the purchase or sale of any security, financial instrument, financial product or service, or to be used in any way for evaluating the merits of participating in any transaction, and shall not constitute a solicitation under any jurisdiction or to any person, if such solicitation under such jurisdiction or to such person would be unlawful. © 2025 JPMorgan Chase \& Co. All rights reserved.


\bibliography{iclr2025_conference}
\bibliographystyle{iclr2025_conference}

\end{document}